\documentclass[journal]{IEEEtran}
\usepackage{amsmath,amsfonts}
\usepackage{algorithmic}
\usepackage{algorithm}
\usepackage{array}
\usepackage[caption=false,font=normalsize,labelfont=sf,textfont=sf]{subfig}
\usepackage{textcomp}
\usepackage{stfloats}
\usepackage{url}
\usepackage{verbatim}
\usepackage{graphicx}
\usepackage{cite}
\usepackage{amsthm}
\usepackage{wrapfig}

\theoremstyle{definition}

\begin{document}
%
\title{Federated TrustChain: Blockchain-Enhanced LLM Training and Unlearning}


%
%
%

\author{Xuhan Zuo, Minghao Wang, Tianqing Zhu$^*$, Lefeng Zhang, Dayong Ye, Shui Yu,~\IEEEmembership{Fellow,~IEEE,} Wanlei Zhou,~\IEEEmembership{Senior Membership,~IEEE}

\thanks{$^*$Tianqing Zhu is the corresponding author with Faculty of Data Science, City University of Macau, Macao (E-mail: tqzhu@cityu.edu.mo)}
\thanks{Xuhan Zuo, Dayong Ye and Shui Yu are with School of Computer Science, University of Technology Sydney, Ultimo 2007, Australia (E-mail: Xuhan.Zuo-1@student.uts.edu.au; Dayong.Ye@uts.edu.au; Shui.Yu@uts.edu.au)}
\thanks{Minghao Wang, Lefeng Zhang and Wanlei Zhou are with the Faculty of Data Science, City University of Macau, Macao (E-mail: sydminghao@gmail.com; lfzhang@cityu.edu.mo; wlzhou@cityu.edu.mo)}
}


%
%

\markboth{Journal of \LaTeX\ Class Files,~Vol.~14, No.~8, August~2015}%
{Shell \MakeLowercase{\textit{et al.}}: Bare Demo of IEEEtran.cls for IEEE Journals}
%



\maketitle

\begin{abstract}

The development of Large Language Models (LLMs) faces a significant challenge: the exhausting of publicly available fresh data. This is because training a LLM needs a large demanding of new data. Federated learning emerges as a promising solution, enabling collaborative model to contribute their private data to LLM global model. However, integrating federated learning with LLMs introduces new challenges, including the lack of transparency and the need for effective unlearning mechanisms. Transparency is essential to ensuring trust and fairness among participants, while accountability is crucial for deterring malicious behaviour and enabling corrective actions when necessary. 
To address these challenges, we propose a novel blockchain-based federated learning framework for LLMs that enhances transparency, accountability, and unlearning capabilities. Our framework leverages blockchain technology to create a tamper-proof record of each model's contributions and introduces an innovative unlearning function that seamlessly integrates with the federated learning mechanism. We investigate the impact of Low-Rank Adaptation (LoRA) hyperparameters on unlearning performance and integrate Hyperledger Fabric to ensure the security, transparency, and verifiability of the unlearning process.
Through comprehensive experiments and analysis, we showcase the effectiveness of our proposed framework in achieving highly effective unlearning in LLMs trained using federated learning. Our findings highlight the feasibility of integrating blockchain technology into federated learning frameworks for LLMs.
\end{abstract}


%
\IEEEpeerreviewmaketitle

\section{Introduction}

The evolution of Large Language Models (LLMs) marks the beginning of a new era in artificial intelligence, significantly altering how we interact with and utilize machine learning \cite{chang2023survey, kasneci2023chatgpt}. As these models progress, a significant challenge becomes apparent: by 2030, publicly available data sources are expected to be insufficient to support the continued growth and development of LLMs \cite{villalobos2022will}. Therefore, the use of private data becomes crucial, not only to sustain development but also as an essential resource for LLMs to access.

With this demand, a significant challenge persists: data owners, aware of the value of LLMs, are hesitant to share their private data because of privacy concerns. At present, individuals have the option to download models and train them on their own private datasets. This method, however, leads to the development of isolated models. These models lack synergy and do not benefit from interconnected learning among various LLMs, underscoring the need for a more cohesive strategy to efficiently utilize private data.

Federated learning emerges as a prominent solution to address the pressing requirement for private data to enhance LLMs \cite{fan2023fate}. This method of collaborative machine learning enables the training of a model on multiple decentralized devices or servers, each of which holds a portion of the entire dataset \cite{zhang2023fedrecovery}. This approach guarantees that confidential information remains on the owner's device, eliminating the need to distribute or consolidate data, thus directly addressing privacy concerns.

However, merging federated learning with LLMs presents a series of new challenges. One major concern is the lack of transparency in the federated learning process when combined with LLMs. The decentralized nature of federated learning makes it difficult to track and verify the contributions of each participating model, as well as to ensure that the collective learning process is not negatively impacted by suboptimal or compromised models. Additionally, the need for effective unlearning mechanisms becomes crucial in this context, as data owners may wish to remove their data from the training process while minimizing the impact on other participants \cite{chang2024class}.

To address these challenges and enhance the transparency and accountability of federated learning in LLM training, we propose the integration of blockchain technology. Blockchain's immutable and distributed ledger provides a secure and transparent record of all transactions and interactions within the federated learning process \cite{wang2020security}. By leveraging blockchain, we can create a tamper-proof record of each model's contributions, facilitating the identification and removal of problematic models without disrupting the overall learning process. 

Furthermore, blockchain enables the implementation of effective unlearning mechanisms, ensuring that data owners can remove their data from the training process while maintaining the integrity of the collective model. Through these dedicated efforts, we introduce an innovative solution that utilizes blockchain technology's strengths to overcome the intricate challenges of training LLMs with private data within a federated learning framework. Our approach represents a substantial step forward in achieving a secure, efficient, and transparent methodology for integrating private data into LLM development.

In addressing the challenges previously outlined, our work offers three significant contributions, each targeting a key aspect of merging federated learning with LLMs via blockchain technology:

\begin{itemize}

    \item We present a blockchain-based architecture meticulously documenting every facet of the federated learning training process. This architecture is crucial for facilitating effective unlearning, as it provides a detailed and unchangeable record of all training actions, ensuring transparency and verifiability at every step.

    \item We introduce an unlearning function within this blockchain environment. This feature is designed to seamlessly integrate with the federated learning mechanism, enabling the targeted removal of specific models or data while preserving the integrity of the wider learning system. Its deployment is vital for upholding the federated learning framework's integrity and effectiveness, allowing it to dynamically respond to changing data privacy requirements.

    \item Our approach strengthens the accountability and verification process by methodically recording unlearning actions on the blockchain. This procedure is essential for evaluating the unlearning process's success.

\end{itemize}

The structure of the paper is as follows: Section II provides a comprehensive review of the existing literature on federated LLMs, unlearning with LLMs, and blockchain's role in enhancing LLMs. Section III lays out fundamental concepts crucial to our discussion, including federated learning, LLMs, LoRA Finetuning, and blockchain technology. Section IV defines the problem and outlines the system model, preparing the groundwork for Section V, which unveils our blockchain-based framework for federated learning. Section VI delves into the privacy and security evaluations of our framework, whereas Section VII measures its overall effectiveness. Finally, Section VIII wraps up the paper by summarizing our key findings and proposing avenues for future investigation.

\section{Related Work}

Large Language Models (LLMs) mark a significant breakthrough in natural language processing (NLP), distinguished by their capability to comprehend, interpret, and produce text that closely mimics human language \cite{min2023recent}. Prominent examples of these models include GPT (Generative Pre-trained Transformer) \cite{floridi2020gpt} and BERT (Bidirectional Encoder Representations from Transformers) \cite{devlin2018bert}, which are trained on vast collections of textual data. This extensive training process equips LLMs with a profound understanding of linguistic subtleties, empowering them to support a broad spectrum of applications. These range from enhancing text completion functionalities to powering complex question-answering systems.

To fully grasp the current research landscape in integrating Large Language Models (LLMs) with federated learning and blockchain technology, we review three key areas: federated learning with LLMs, unlearning mechanisms in LLMs, and the application of blockchain to LLMs. Federated learning allows training LLMs on decentralized datasets while preserving privacy, but introduces challenges like ensuring model integrity and enabling efficient unlearning. Unlearning is crucial for maintaining data privacy and regulatory compliance. Blockchain technology can potentially enhance the security, transparency, and verifiability of federated learning and unlearning in LLMs. Reviewing these interconnected areas helps identify state-of-the-art approaches, limitations, and opportunities for synergistic integration.


\subsection{Federated LLM}

In addressing the exhaustion of public data resources for LLM training, federated learning emerges as a potent solution. By enabling multiple participants to collaboratively train a model without sharing their raw data, federated LLM can access a wider array of diverse and representative datasets. 

Chen et al. conclude the concept of federated Large-Scale Language Models (LLMs), which includes federated pre-training, fine-tuning, and prompt engineering, and explore the unique challenges and potential engineering strategies within this framework, highlighting its advantages over traditional LLM training approaches\cite{chen2023federated}. In Gupta et al. study\cite{gupta2022recovering}, they introduce FILM, a novel attack methodology for federated learning of language models. They demonstrate for the first time the feasibility of recovering text data from large batch sizes and evaluating various defence strategies, thereby suggesting new directions for enhancing privacy in language model training. This paper \cite{fan2023fate} proposed an industrial federated learning framework, which is designed to facilitate the efficient training of large language models. This framework addressed the dual challenges of computational resources and data privacy. The LP-FL methodology prioritizes the reduction of model parameters within the federated learning framework \cite{jiang2023low}, this method employs Low-Rank Adaptation (LoRA) technology to construct compact learnable parameters, enabling effective local model fine-tuning and sustainable global model federation. FederatedScope-LLM (FS-LLM) provides a robust framework for optimizing LLM in a network. FS-LLM processes from data preparation to outcome assessment and facilitating diverse computational strategies \cite{kuang2023federatedscope}. Therefore, there are limitations among these frameworks due to the lack of transparency in the LLM training processes. \cite{zhao2024enhancing} proposed an automated data quality control pipeline for federated fine-tuning of LLM, by utilizing data valuation algorithms, this pipeline assesses the quality of training samples across collaborative platforms, thereby enhancing model performance while preserving data privacy. There are also research concerns in the wireless field, \cite{jiang2024personalized} addresses significant challenges, including privacy concerns, inefficient data handling, and high communication costs, and demonstrates the effectiveness of these methods through simulations. In the Ro et al. research, they demonstrate that scale-invariant modifications to the Coupled Input Forget Gate (CIFG) and transformer models significantly enhance federated learning performance by improving convergence speeds and offering an improved privacy-utility trade-off \cite{ro2024efficient}.

\subsection{Unlearning with LLM}

The rapid advancements in large language models (LLMs) have led to remarkable breakthroughs in natural language processing and artificial intelligence. However, as these models are trained on vast amounts of data, they may inadvertently learn and perpetuate undesirable behaviors, biases, and harmful information. To address this issue, researchers have recently turned their attention to the concept of unlearning in LLMs.

In \cite{yao2023large} paper, the authors explore the novel concept of unlearning in large language models (LLMs). They present a method that utilizes only negative examples to efficiently remove undesirable behaviors, demonstrating its effectiveness in alignment while significantly reducing computational resources compared to traditional reinforcement learning from human feedback (RLHF). While there is another paper introduces a data-driven unlearning approach for large language models (LLMs), utilizing a fine-tuning method informed by the importance of weights and relabeling during the pre-training phase of LLMs \cite{yu2023unlearning}. This method adjusts word embedding, involving identifying and neutralizing bias vectors within the embedding space to prevent biased associations. Wang et al.\cite{wang2023kga} proposed an unlearning framework called Knowledge Gap Alignment (KGA), emphasizing its capability to efficiently handle large-scale data removal requests with significant accuracy. However, the inability of KGA to guarantee the complete removal of data influences also faces the challenge of maintaining extra data sets and models. Si et al.\cite{si2023knowledge} explores the technical challenges of knowledge unlearning in large language models (LLMs), specifically introducing parameter optimization, parameter merging, and in-context learning as methods to efficiently remove harmful or biased data while maintaining the integrity of the models. This approach not only advances the field of responsible AI but also opens new avenues for enhancing data privacy and model impartiality. Huang et al. claim an innovation offset unlearning framework tailored for the black box LLM\cite{huang2024offset}. This framework effectively addresses the challenge of unlearning problematic training data in LLMs without requiring access to internal model weight, thus offering a versatile solution for adapting current unlearning algorithms.

\subsection{Blockchain with LLM}
Blockchain technology and artificial intelligence (AI) have emerged as two of the most transformative technologies of our time. The integration of these technologies has the potential to revolutionize various industries and address critical challenges faced by both domains. Recent research has explored the synergistic relationship between blockchain and AI, particularly focusing on the integration of blockchain with large language models (LLMs) and generative AI (GAI) techniques.

Luo et al.\cite{luo2023bc4llm} introduce the concept of "Blockchain for LLM" (BC4LLM), which aims to empower LLMs with the superior security features of blockchain technology, enabling reliable learning corpora, secure training processes, and identifiable generated content. This paper presents emerging solutions that showcase the effectiveness of GAI in detecting unknown blockchain attacks and smart contract vulnerabilities, designing key secret sharing schemes, and enhancing privacy. Through a case study, they demonstrate that the generative diffusion model, a GAI approach, can significantly optimize blockchain network performance metrics, outperforming traditional AI approaches in terms of convergence speed, rewards, throughput, and latency \cite{nguyen2024generative}. Mboma et al. propose a novel approach to combat academic document fraud by integrating Large Language Models (LLMs), specifically the Bidirectional Encoder Representations from Transformers (BERT), with blockchain and Interplanetary File System (IPFS) technologies to pre-validate academic documents before certification \cite{mboma2024integrating}. LLMChain, a decentralized blockchain-based reputation system, assists users and entities in identifying the most trustworthy LLM for their specific needs while providing valuable information to LLM developers for model refinement \cite{bouchiha2024llmchain}. This framework demonstrated through evaluation across two benchmark datasets, making it a significant contribution to the field of trustworthy and transparent LLM assessment.

\subsection{Conclusion}
Despite progress in federated learning, unlearning, and blockchain integration with LLMs, several common limitations persist:
\begin{itemize}
    \item Lack of comprehensive frameworks that integrate these approaches for enhanced security and transparency.
    \item Limited scalability and efficiency of current unlearning mechanisms in large-scale federated learning settings.
    \item Insufficient privacy and security guarantees in federated learning, with potential for attacks or information leakage.
    \item Absence of standardized frameworks and protocols for integrating these technologies, hindering interoperability and adoption.

\end{itemize}

Addressing these limitations requires developing a comprehensive framework that integrates federated learning, efficient unlearning, and blockchain technology to enable secure, transparent, and privacy-preserving LLM training on decentralized datasets.

\section{Preliminary}

\subsection{Federated Learning}

Federated Learning (FL) is a distributed machine learning approach that allows multiple devices or servers, each possessing its own local data samples, to collaboratively develop a model without the need to share their data directly \cite{wang2023blockchain}. This concept can be mathematically represented as:
\begin{equation}
    FL = \{D_1, D_2, \ldots, D_n\}
\end{equation}

where \(D_i\) denotes the local dataset of the \(i\)-th participant in the federation, and \(n\) represents the total number of participants. The core aim of FL is to build a comprehensive global model \(G\) that assimilates the knowledge from all local datasets, thereby improving model efficacy and ensuring data privacy.

The federated learning training protocol unfolds through several essential steps:
\begin{enumerate}
\item Initially, a global model \(G\) is distributed among all participants.
\item Each participant \(i\) refines this global model using their own data \(D_i\), resulting in an updated local model \(M_i\).
\item These updated local models \(M_i\) are then consolidated to refine the global model \(G\), utilizing secure aggregation techniques to protect the privacy of individual updates.
\end{enumerate}

The process undergoes multiple iterations, with the global model \(G\) being incrementally improved in each round. The aggregation function, often employing a form of weighted average, is pivotal in merging the local updates into a cohesive global model. This can be represented mathematically as:
\begin{equation}
G = Agg(M_1, M_2, \ldots, M_n)
\end{equation}
where \(Agg\) denotes the aggregation mechanism used to combine the updates.

\subsection{Large Language Models (LLMs)}

The architecture of LLMs is fundamentally based on transformer models, characterized by a series of layers that systematically process the input text data \cite{yang2023harnessing}. At the heart of these models is the self-attention mechanism, a crucial feature that enables LLMs to assess the significance of each word in a sentence, thereby crafting responses that are contextually coherent \cite{caballero2023exploring}. The mathematical representation of an LLM's output can be succinctly expressed as:
\begin{equation}
O = F(I; \theta)
\end{equation}

The equation represents the relationship where \(I\) is the input text, 
\(O\) the output generated by the model, \(F\) the function embodied by the LLM, and \(\theta\) the set of parameters honed during training. 
The training process for LLMs involves fine-tuning these parameters (\(\theta\)) to reduce the discrepancy between the model's output and the expected output, enhancing the model's precision in generating relevant responses. The sheer size of LLMs, with potentially billions of parameters, endows them with exceptional levels of language comprehension and production.

However, deploying LLMs is not without its hurdles. The need for extensive datasets for training and considerable computational power are significant barriers \cite{hadi2023survey}. Furthermore, incorporating LLMs into federated learning environments brings extra challenges, including preserving model performance and privacy across decentralized data sources.

\subsection{LoRA Fine-tuning}

LoRA (Low-Rank Adaptation) \cite{hu2021lora} offers an innovative method for fine-tuning Large Language Models (LLMs) that balances efficiency with effectiveness, especially valuable in scenarios demanding model adaptability and computational thrift. Unlike traditional approaches that modify the original model parameters, LoRA adapts pre-trained LLMs to specific tasks through a low-rank decomposition technique. This method introduces additional trainable parameters, enabling the model to undergo task-specific adjustments without direct changes to its foundational parameters.

LoRA is particularly well-suited for federated learning environments, as it allows for efficient and targeted fine-tuning of LLMs across multiple participants without the need to share the entire model. Additionally, LoRA's low-rank decomposition approach makes it an ideal candidate for enabling effective unlearning mechanisms, as it allows for the selective modification of specific model components without affecting the overall model performance.

The core of LoRA's innovation lies in its approach to modifying the attention and feed-forward layers of transformer-based Large Language Models (LLMs) by integrating low-rank matrices. Specifically, for a weight matrix $W \in \mathbb{R}^{m \times n}$ within a transformer layer, LoRA introduces two smaller matrices, $A \in \mathbb{R}^{m \times k}$ and $B \in \mathbb{R}^{k \times n}$, where $k \ll \min(m,n)$. The adaptation of the original weight matrix $W$ can be mathematically described as:
\begin{equation}
    W' = W + AB
\end{equation}
where $W'$ denotes the adapted weight matrix, and $AB$ is the low-rank update applied to $W$.

Such a strategy enables substantial customization of the model with only a modest increase in parameters, maintaining the extensive knowledge of the pre-trained LLM while introducing task-specific adjustments efficiently. During the fine-tuning phase, only the low-rank matrices $A$ and $B$ are updated, significantly lowering the computational demands typically seen with large-scale model training. Consequently, LoRA's approach to fine-tuning offers a scalable, resource-effective method for tailoring LLMs to diverse tasks and sectors. This is especially advantageous in federated learning settings, where computational efficiency and the flexibility to adapt to various tasks are paramount.

\subsection{Blockchain}

Blockchain technology is a decentralized ledger system that offers a secure and transparent method for recording transactions across multiple computers \cite{wang2023differentially}. Its foundation relies on cryptography principles, ensuring that each entry in the ledger is immutable and verifiable \cite{sunyaev2020distributed}. While this technology is the backbone of cryptocurrencies like Bitcoin and Ethereum, it also extends its applications to secure transactional data in various sectors, including supply chain management, healthcare, and, as explored in this paper, federated learning environments.

A blockchain comprises a sequence of blocks, each containing a list of transactions. These blocks are interconnected through a cryptographic hash, linking each block to its predecessor and forming a chain. This structure is mathematically represented as \(B_1 \rightarrow B_2 \rightarrow \ldots \rightarrow B_n\), where \(B_i\) symbolizes the \(i\)-th block in the chain, and \(n\) represents the total number of blocks. The integrity of the chain is preserved by consensus algorithms, which ensure that all instances of the distributed ledger are synchronized and in agreement on the transaction sequence.


\section{Problem Definition and System Model}

\subsection{Problem Definition}

The integration of Large Language Models (LLMs) with federated learning, supported by a blockchain framework, introduces distinct aims that require a precise problem definition. These aims arise from the complexities of managing private data, ensuring model integrity, and implementing efficient unlearning processes. We formalize these aims as follows:

\begin{enumerate}
    \item \textbf{Data Privacy and Model Efficacy:} Federated learning aims to train LLMs on a collection of private datasets (\(D_c\)) across various clients (\(C_{id}\)) without breaching data privacy. The main challenge is to enhance the global LLM (\(LLM_g\)) performance while respecting privacy constraints, posing an optimization problem of maximizing \(LLM_g\)'s efficacy across the federated network without direct access to \(D_c\).
    
    \item \textbf{Model Integrity and Security:} Within federated learning, each client boosts the global model by updating parameters using their local data. This decentralized method, however, exposes vulnerabilities like the potential for backdoor attacks or model tampering. It is crucial to secure the global model (\(LLM_g\)) and the aggregation process (\(A_{id}\), \(JWT\)), especially when model updates come from possibly unreliable sources.
    
    \item \textbf{Efficient Unlearning Mechanisms:} The changing dynamics of data privacy laws and data itself demand an effective mechanism for removing specific data (\(D_{forget}\)) from the trained model (\(LLM_g\)). The challenge lies in developing a process that allows \(LLM_g\) to selectively discard \(D_{forget}\) through unlearning epochs (\(E_u\)) and LoRA parameters (\(\lambda\)), with minimal detriment to the model's overall performance.
    
    \item \textbf{Immutable Record Keeping and Verification:} The decentralized nature of federated learning complicates the monitoring and validation of model updates, contributions, and unlearning activities. It is vital to establish a transparent and unchangeable record-keeping system on a blockchain (\(SC\), \(T_{id}\)) that logs all actions related to model training, updating, and unlearning. This system must support the authentication of actions (\(parameters\), \(D_{validate}\)) to maintain integrity and accountability in the federated learning process.
\end{enumerate}

Our proposed blockchain-based framework seeks to achieve these aims by employing cryptography techniques (\(JWT\), \(P_k\), \(S_k\)) for secure client registration, ensuring model integrity through a secure aggregation process, enabling efficient unlearning, and maintaining an immutable ledger for action verification. This strategy aims to improve the privacy, security, and effectiveness of LLM development within a federated learning framework.

\subsection{System Model}
Our system model achieve these aims by integrating Large Language Models (LLMs) with federated learning, underpinned by the security and immutability of blockchain technology. The model encompasses the processes of client registration, federated learning training, model aggregation, and the unlearning process, each facilitated by smart contracts (SC) on a blockchain network. Below, we detail the components and their interactions within the system.

\subsubsection{Participants}
The system includes several types of participants, each playing a pivotal role in the federated learning ecosystem:
\begin{itemize}
    \item \textbf{Clients} (\(C_{id}\)): Entities with private datasets (\(D_c\)) looking to contribute to and benefit from the global LLM (\(LLM_g\)) without sacrificing data privacy.
    \item \textbf{Agents} (\(A_{id}\)): Individuals responsible for managing the aggregation of local model updates into the global model and facilitating the unlearning process. Agents operate with verification and authorization provided by JWTs, ensuring secure interactions.
    \item \textbf{Smart Contracts} (SC): Autonomous programs on the blockchain executing predefined operations such as client registration, model aggregation, and the execution of the unlearning process, thereby ensuring transparency, security, and trust.
\end{itemize}

\subsubsection{Process Flow}
The system model revolves around key processes, orchestrated through the interaction of participants:
\begin{itemize}
    \item \textbf{Client Registration:} Clients (\(C_{id}\)) register in the system through a secure process involving the generation of a public-secret key pair (\(P_k, S_k\)) and obtaining a JSON Web Token (JWT) for secure communication. This process guarantees each client's unique identification and secure authentication within the system.
    \item \textbf{Federated Learning with LLM Training:} Clients engage in the federated learning process by locally training the LLM on their private datasets (\(D_c\)) and sharing the learned parameters with the global model (\(LLM_g\)), all while keeping their data confidential. This iterative process across multiple epochs aims to enhance the global model's precision and robustness.
    \item \textbf{Model Aggregation:} Agents (\(A_{id}\)), verified via JWTs, consolidate the parameters from clients into the global model (\(LLM_g\)). Smart contracts (SC) secure and oversee this aggregation process, ensuring only authorized updates enhance the global model.
    \item \textbf{Unlearning Process:} The system facilitates an efficient unlearning mechanism allowing the selective omission of data (\(D_{forget}\)) from the global model (\(LLM_g\)). Utilizing unlearning epochs (\(E_u\)) and specific parameters (\(\lambda\)), the model adjusts without losing learning from other data contributions.
    \item \textbf{Blockchain for Security and Transparency:} All activities, including client registration, model updates, and the unlearning actions, are recorded on the blockchain via smart contracts (SC). This immutable ledger elevates the system's security, transparency, and trust.
\end{itemize}

\section{Proposed Framework}
\subsection{Overview}

Our proposed system introduces a novel framework that seamlessly integrates Large Language Models (LLMs) with federated learning, leveraging the security and transparency provided by blockchain technology. This meticulously designed integration aims to harness the advantages of federated learning for training LLMs on decentralized private datasets while preserving data privacy, ensuring model integrity, and facilitating an efficient unlearning process. 

\begin{figure}
\includegraphics[width=0.5\textwidth]{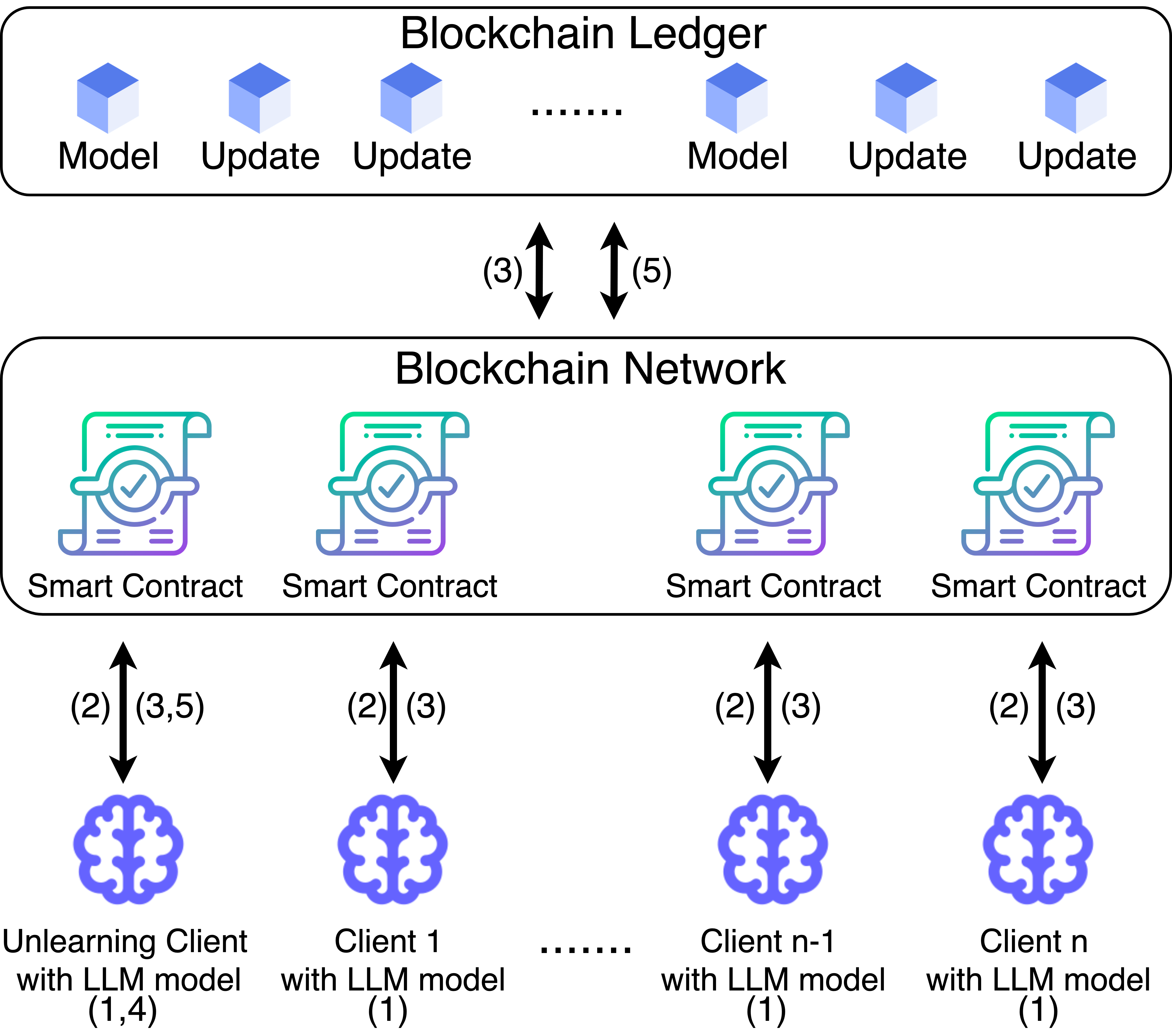}
\caption{Overview and process of our proposed system. (1) Client register. (2) Federated learning LLM training process. (3) Model aggregation process. (4) Unlearning process using LoRA for forgetting. (5) Unlearning verification and submitting unlearning results. }
\label{overview}
\end{figure}

Figure~\ref{overview} presents an overview of our proposed system. To begin, all clients must complete the registration process within the blockchain network. Once registration is finalized, the blockchain network initiates the federated learning training process. The global model is transferred within the blockchain network through a smart contract, followed by the aggregation of the model. In the event that a client wishes to erase their private data, the unlearning process is triggered, employing LoRA to facilitate efficient forgetting. Subsequently, a verification process is conducted to ensure the integrity of the unlearning procedure. Upon successful verification, the system seamlessly returns to the standard federated learning training process. The implementation details of our meticulously designed framework are outlined below.

\subsection{Client Register}
In our proposed framework, every \textit{Client} need to enroll in blockchain network first. Algorithm~\ref{Register} facilitates a straightforward method for registering a client using a unique identifier and securing their communication with a JSON Web Token (JWT). Initially, the process verifies if the client's unique identifier ($C_{id}$) is already present in the user pool ($U_{pool}$). If the identifier exists, the registration halts, indicating the client already exists. Otherwise, the algorithm proceeds to generate a public-secret key pair using the $keyGenerator()$ function. With these keys, it then creates a \textit{JWT} for the client. This \textit{JWT}, along with the client ID, is securely stored, effectively registering the client. The user pool is updated to include the new client ID, marking the registration successful. The algorithm concludes by returning a success status and the generated \textit{JWT}, signifying the client's successful registration and their secure token for future communications. This process ensures a secure registration framework by leveraging cryptographic keys and \textit{JWT}s, ensuring both security and simplicity in client management.
\begin{algorithm}
\caption{Client Register}
\label{Register}
\begin{algorithmic}[1]
\REQUIRE \textit{C$_{id}$, keyGenerator(),generateJWT()}
\ENSURE \textit{RegisterSucess, jwt token}
\STATE \textit{RegisterSucess = False};
\IF{\textit{C$_{id}$ $\in$ U$_{pool}$}}
    \RETURN \textit{C$_{id}$} already existed.
\ENDIF
\STATE ($P_k$, $S_k$) $\xleftarrow{}$\textit{keyGenerator()};
\STATE \textit{jwt} = generateJWT($P_k$, $S_k$);
\STATE \textit{C$_{id}$} $\xleftarrow{}$ $jwt$ $\xleftarrow{}$ SC;

\STATE \( U_{\text{pool}} = U_{\text{pool}} \cup C_{\text{id}} \);
\STATE \textit{RegisterSuccess = True};
\RETURN \textit{RegisterSucess, jwt}
\end{algorithmic}
\end{algorithm}

\subsection{Federated Learning with LLM Training Process}

The federated learning process for training large language models (LLMs) involves multiple clients collaborating to improve a global model without sharing their private data directly. This process ensures data privacy, security, and decentralization. First, an agent initiates the process by sending the LLM to the smart contract (SC). The SC verifies the agent's identity using a JSON Web Token (jwt) and uploads the global model ($LLM_g$) to the blockchain network. The $LLM_g$ is then distributed to the participating clients. During each training epoch, clients perform federated learning on their private datasets to improve the $LLM_g$. After training, the clients send the updated LLM parameters to the SC for verification and aggregation. The SC verifies each client's identity using their jwt tokens and publishes the updated parameters and client information to the blockchain network. This process is repeated for a specified number of epochs. Upon completion, the algorithm returns the status of the LLM upload and training process.

Algorithm~\ref{FLLLM} outlines the steps for training a large language model (LLM) in a federated learning environment, emphasizing security and decentralization.

\begin{algorithm}
\caption{Client LLM Training Process}
\label{FLLLM}
\begin{algorithmic}[1]
\REQUIRE \( C_{\text{id}}, A_{\text{id}}, \text{jwt}, epochs, LLM \)
\ENSURE \textit{UploadSuccess, TrainingProcess}

\STATE \( \textit{UploadSuccess, TrainingProcess} = \text{False} \);
\STATE SC check \textit{Agent's} identity;
\IF{\textit{Agent's jwt token ineligibility}}
    \RETURN \text{Agent jwt token expired}
\ENDIF

\STATE \textit{Agent} sends \( LLM \) to SC;

\STATE \textit{SC} verifies and uploads the global model \( LLM_g \) to the blockchain network;

\STATE \( LLM_g \) = \( LLM \);

\STATE \( \textit{UploadSuccess} = \text{True} \);
\STATE SC send the $LLM_g$ to $Client$;
\FOR{$epoch = 1$ to $n$}
\STATE $Clients$ do the federated learning training process according to their different private dataset $D_c$ for $LLM_g$
\STATE $Clients$ send the \textit{parameters} of $LLM$ to SC
\STATE SC verify the $Client$ identity 
\IF{\textit{Client's jwt token ineligibility}}
    \RETURN \text{Client identity check false}
    \ELSE \STATE SC publish the $parameters$ and $Clients$ information in blockchain network
\ENDIF

\ENDFOR
\STATE \textit{TrainingProcess} = \textit{True};

\RETURN \( \textit{UploadSuccess, TrainingProcess}\)
\end{algorithmic}
\end{algorithm}

The algorithm begins by initializing two boolean variables, \textit{UploadSuccess} and \textit{TrainingProcess}, to False. These variables track the status of the LLM upload and the training process, respectively. The required inputs include the client identifier ($C_{id}$), agent identifier ($A_{id}$), JSON Web Token (jwt) for authentication, number of training epochs, and the LLM to be trained.

The SC first verifies the agent's identity using the provided jwt. If the token is invalid or has expired, the process is terminated, and an error message is returned. Upon successful authentication, the agent sends the LLM to the SC, which then verifies and uploads the global model ($LLM_g$) to the blockchain network, ensuring the model's integrity and security in a decentralized environment. The $LLM_g$ is initialized with the agent's LLM, and the \textit{UploadSuccess} variable is set to True, indicating the successful upload of the model.

The SC then distributes the $LLM_g$ to the participating clients for training. The training process is conducted iteratively for a specified number of epochs ($n$). During each epoch, clients perform federated learning on their private datasets ($D_c$) to improve the $LLM_g$. After training, the clients send the updated LLM parameters to the SC for verification and aggregation.

The SC verifies each client's identity using their jwt tokens. If a client's token is invalid, the process is terminated for that client, and an error message is returned. Otherwise, the SC publishes the updated parameters and client information to the blockchain network, ensuring transparency and security.
Upon completing the specified number of training epochs, the \textit{TrainingProcess} variable is set to True, indicating the successful completion of the federated learning process. Finally, the algorithm returns the values of \textit{UploadSuccess} and \textit{TrainingProcess}, providing information about the status of the LLM upload and training process.

\subsection{Model Aggregation Process}

The model aggregation process is a crucial step in updating the global language model ($LLM_g$) in a secure and decentralized manner. This process is initiated by an agent who requests the latest model parameters from the blockchain network. The smart contract (SC) verifies the agent's identity using a JSON Web Token (JWT). Upon successful authentication, the SC sends the parameters to the agent, who then updates the $LLM$ and generates a new model version ($LLM_n$). The agent sends $LLM_n$ back to the SC, which uploads it to the blockchain network, ensuring a secure and transparent record of the update. Finally, $LLM_g$ is updated to reflect the changes in $LLM_n$, completing the model aggregation process.

Algorithm~\ref{Aggregation} outlines the procedure for aggregating updates to a large language model ($LLM$) in a secure and decentralized manner, leveraging a blockchain network for data integrity and transparency.

\begin{algorithm}
\caption{Model Aggregation Process}
\label{Aggregation}
\begin{algorithmic}[1]
\REQUIRE \textit{$A_{id}$} \textit{JWT}, \textit{$parameters$}

\ENSURE \textit{ModelAggregation, $LLM_g$}

\STATE \textit{ModelAggregation} = False;

\STATE \textit{Agent} wants to get \textit{parameters} from blockchain network;
\STATE SC check the \textit{Agent} identity;
\IF{\textit{Agent's jwt token ineligibility}}
    \RETURN \textit{Agent} identity check false
    \ELSE 
    \STATE SC send \textit{parameters} to \textit{Agent}
    \ENDIF

\STATE \textit{Agent} updating \textit{LLM} according to \textit{parameters} and generating new model \textit{$LLM_n$};
\STATE Agent send the new model \textit{$LLM_n$} to SC;
\STATE SC upload the \textit{$LLM_n$} to blockchain network;
\STATE \textit{$LLM_g$} $\xleftarrow{}$ \textit{$LLM_n$} ;

\STATE \textit{ModelAggregation = True};
\RETURN \textit{ModelAggregation, $LLM_g$}
\end{algorithmic}
\end{algorithm}

The process begins by initializing the $ModelAggregation$ flag to False, indicating that the aggregation process has not yet started. An agent, identified by $A_{id}$ and authenticated using a $JWT$, requests the latest model parameters from the blockchain network. These parameters will be used to update the $LLM$ to a new version, $LLM_n$.

The SC verifies the agent's identity by checking the validity of the provided $JWT$. If the $JWT$ is invalid, the process is terminated, and the agent is informed that their identity check has failed. This step ensures that only authorized agents can retrieve and update model parameters, maintaining the system's security.

If the agent's identity is successfully verified, the SC sends the requested parameters to the agent. The agent then uses these parameters to update the $LLM$, generating a new model version, $LLM_n$. This step involves applying the aggregated updates from various sources to improve the model's performance or capabilities based on newly acquired data or insights.

After generating $LLM_n$, the agent sends this new model version back to the SC. The SC uploads $LLM_n$ to the blockchain network, ensuring that the update is securely and transparently recorded. The global version of the $LLM$, $LLM_g$, is then updated to reflect the changes in $LLM_n$, completing the model update process.

Finally, the $ModelAggregation$ flag is set to True, indicating the successful aggregation of the model updates. The algorithm returns this flag along with $LLM_g$, the updated global model, signifying the end of the aggregation process.

\subsection{Unlearning Process}

The unlearning process is a crucial step in selectively forgetting specific data from a large language model (LLM) due to data sensitivity or correction needs. This process begins with the initialization of a local version of the LLM ($LLM_{local}$) using the parameters of the global model ($LLM_g$). An adapter ($A$) is then constructed within $LLM_{local}$ to facilitate the forgetting of the specified dataset ($D_{forget}$). The core of the unlearning process involves several epochs of training, where a forward pass of $D_{forget}$ is performed through $LLM_{local}$ to identify the features associated with the data points that need to be forgotten. Gradients are then computed for $LLM_{local}$, emphasizing the data to be unlearned. The Low-Rank Adaptation (LoRA) technique is applied to the adapter's gradients to focus the unlearning process on the identified features. Finally, $LLM_{local}$'s parameters are updated using the adjusted gradients and a specified learning rate, gradually leading to the forgetting of the specified data points. The algorithm returns the updated parameters, representing the outcome of the forgetting process.

Algorithm~\ref{unlearning} describes the procedure for selectively forgetting specific data from a large language model (LLM).

\begin{algorithm}
\caption{Unlearning Process using LoRA for Forgetting}
\label{unlearning}
\begin{algorithmic}[1]

\REQUIRE $LLM_g$, $D_{forget}$ (Dataset to forget), Learning rate $\eta$, Unlearning epochs $E_u$, LoRA parameters $\lambda$
\ENSURE \textit{parameters}
\STATE Unlearning Request due to data sensitivity or correction needs;
\STATE Initialize unlearning model $LLM_{local}$ with $LLM_g$
\STATE \textit{Adapter} $A$ constructed for $LLM_{local}$ targeting forgetting process

\FOR{$epoch = 1$ to $E_u$}
    \STATE Perform a forward pass with $D_{forget}$ through $LLM_{local}$ to identify features to forget
    \STATE Compute gradients for $LLM_{local}$ emphasizing data points in $D_{forget}$ to be forgotten
    \STATE Apply LoRA to adjust gradients of adapter $A$ using parameters $\lambda$, focusing on unlearning
    \STATE Update $LLM_{local}$'s parameters using the adjusted gradients and learning rate $\eta$, facilitating forgetting
\ENDFOR

\STATE Calculate the updating \textit{parameters} indicative of the forgetting process between $LLM_{local}$ and $LLM_g$;

\RETURN \textit{parameters}

\end{algorithmic}
\end{algorithm}

The process begins with the need to remove certain data points from a global language learning model ($LLM_g$) due to their sensitivity or incorrectness. To achieve this, a local version of the LLM, denoted as $LLM_{local}$, is initialized with the parameters of $LLM_g$. An adapter, $A$, is then constructed within $LLM_{local}$ specifically designed to target and facilitate the forgetting of the specified dataset, $D_{forget}$.

The core of the unlearning process involves several epochs of training, defined by the parameter $E_u$. In each epoch, the algorithm performs a forward pass of $D_{forget}$ through $LLM_{local}$ to identify the features associated with the data points that need to be forgotten. Following this, gradients are computed for $LLM_{local}$ with an emphasis on the data to be unlearned, highlighting what needs to be forgotten.

The LoRA technique is applied to the adapter $A$'s gradients using parameters $\lambda$. LoRA is instrumental in focusing the unlearning process by adjusting the gradients to specifically target the forgetting of the identified features. With these adjusted gradients, $LLM_{local}$'s parameters are updated using the specified learning rate $\eta$. This iterative process of adjustment and updating gradually leads to the forgetting of the specified data points from $D_{forget}$.

Upon completion of the unlearning epochs, the algorithm calculates the parameters that indicate the changes made to $LLM_{local}$ in comparison to $LLM_g$. These parameters represent the outcome of the forgetting process, effectively capturing the essence of what has been unlearned.

The algorithm concludes by returning these updated parameters, signifying the successful exclusion of sensitive or incorrect data from the language model. Through this structured process, the algorithm ensures that the unlearning is specific, efficient, and aligned with the requirements of data sensitivity or correction, thereby maintaining the integrity and relevance of the LLM.

\subsection{Unlearning Verification and Submitting Unlearning Results}

The unlearning verification and submission process is a critical step in ensuring the integrity and transparency of the unlearning results in a large language model. The process begins with the client sending the updated parameters, resulting from an unlearning process, to the smart contract (SC). The SC validates the client's credentials through their JSON Web Token (JWT). If the client's identity is successfully verified, the SC initializes an updated version of the language learning model ($LLM_{updated}$) with the new parameters. The SC then employs a validation dataset ($D_{validate}$) to assess the efficacy of the unlearning process by calculating the training loss and accuracy of $LLM_{updated}$. If the unlearning results satisfy predefined verification criteria, the SC submits the updated parameters to a blockchain network. An agent downloads these parameters from the blockchain for weight integration into the global model. The SC records the updated model's weights on the blockchain, ensuring transparency and traceability. Additionally, the SC logs a Transaction ID ($T_{id}$), providing verifiable proof of submission and an integration request. The process concludes with the return of the Transaction ID, signifying the successful verification and submission of the unlearning results.

Algorithm~\ref{Verification} details the steps for verifying the results of an unlearning process in a large language model and subsequently submitting these results for integration and transparency.
\begin{algorithm}
\caption{Unlearning Verification and Submitting Unlearning Results}
\label{Verification}
\begin{algorithmic}[1]

\REQUIRE \textit{parameters}, Validation dataset $D_{validate}$, \textit{Client}
\ENSURE \textit{parameters}

\STATE \textit{Client} send the \textit{parameters} to SC;
\IF{\textit{Client's jwt token ineligibility}}
    \RETURN \text{Client identity check false}
\ENDIF
\STATE SC instantiate the updated language learning model $LLM_{updated}$ with the received parameters;
\STATE SC use the validation dataset $D_{validate}$ to evaluate $LLM_{updated}$. Calculate the training loss and accuracy to measure the impact of the unlearning process.
\IF{\textit{Verification criteria are met}}
    \STATE SC send the \textit{parameters} to blockchain networks
    \STATE $Agent$ downloads \textit{parameters} from blockchain network for weight integration. 
    \STATE SC ensuring that the updated weights are recorded on the blockchain, providing transparency and traceability
    \STATE SC record the Transaction ID $T_{id}$, which serves as proof of submission and integration request, facilitating tracking and verification in the blockchain ledger.
\ENDIF
\STATE Continue for future federated learning process;

\RETURN $T_{id}$

\end{algorithmic}
\end{algorithm}

The process commences with the client sending the updated parameters, resulting from an unlearning process, to the SC. These parameters are intended to modify a language learning model by excluding specific, potentially sensitive, or incorrect data. Initially, the client's credentials are validated through their JWT token. If the token does not pass the eligibility check, the process halts, indicating a failure in client identity verification.

Assuming successful verification, the SC then initializes an updated version of the language learning model ($LLM_{updated}$) with the new parameters. The SC employs a validation dataset ($D_{validate}$) to assess the efficacy of the unlearning process. This assessment involves calculating the training loss and accuracy of $LLM_{updated}$ to gauge the impact of the modifications.

If the unlearning results satisfy predefined verification criteria, which indicate that the data has been effectively forgotten without compromising the model's overall performance, the SC will submit the updated parameters to a blockchain network. This submission is not merely for record-keeping; an agent then downloads these parameters from the blockchain for weight integration into the global model.

Recording the updated model's weights on the blockchain ensures that the unlearning process is transparent and traceable. Furthermore, the SC logs a Transaction ID ($T_{id}$), providing verifiable proof of submission and an integration request. This ID facilitates tracking and verification within the blockchain ledger, offering a transparent audit trail of the changes made to the language model.

The process culminates with the return of the Transaction ID, signifying the successful verification and submission of the unlearning results. This structured approach not only secures the integrity of the model by removing unwanted data but also enhances accountability and transparency through blockchain technology.

\section{Privacy and Security Analysis}
Our proposed blockchain-based federated learning framework with unlearning capabilities for Large Language Models (LLMs) is designed to address critical privacy and security challenges. By leveraging the inherent features of federated learning, blockchain technology, and efficient unlearning mechanisms, our approach provides a comprehensive solution for secure and privacy-preserving LLM training.

\subsection{Privacy Analysis}

Federated learning, a core component of our framework, enables the training of LLMs across multiple participants without the need for direct data sharing. This decentralized approach ensures that sensitive data remains within the control of each participant, minimizing the risk of data breaches and unauthorized access.

From a theoretical perspective, federated learning can be modeled as an optimization problem that aims to minimize the global loss function while keeping the data locally \cite{yang2019federated}. This can be represented as:
\begin{equation}
\min_{w} F(w) = \sum_{i=1}^{k} p_i F_i(w)
\end{equation}

where $w$ is the global model parameters, $F(w)$ is the global loss function, $F_i(w)$ is the local loss function of the $i$-th participant, $p_i$ is the weight of the $i$-th participant, and $k$ is the total number of participants.

By minimizing the global loss function, federated learning enables the optimization of the global model without directly sharing raw data, leveraging the data distributed across local participants while protecting privacy and improving model performance.

Our framework further enhances privacy protection by integrating blockchain technology, which provides a secure and immutable record of all transactions and interactions within the federated learning process. The use of smart contracts in our framework automates the execution of predefined rules and conditions, ensuring that all participants adhere to agreed-upon privacy policies. This automation minimizes the potential for human error and reduces the risk of unauthorized data access or manipulation.

Moreover, blockchain technology can provide privacy protection for the federated learning process \cite{lu2019blockchain}. By leveraging the immutability and distributed consensus mechanisms of blockchain, it ensures that all participants follow predefined privacy policies and prevents malicious behavior. In our framework, smart contracts automatically enforce these policies, further reducing the risks of human error and unauthorized data access.

The unlearning mechanism embedded in our framework allows for the selective removal of specific data points or model updates, enabling participants to maintain control over their data and comply with evolving privacy regulations. The unlearning process can be theoretically formulated as a constrained optimization problem \cite{ginart2019making}, where the objective is to minimize the impact of the removed data on the model's performance while satisfying the unlearning constraints:
\begin{equation}
\min_{w} F(w) = \sum_{i=1}^{k} p_i F_i(w)
\end{equation}
\begin{equation*}
\text{s.t.} \quad w \in W_u
\end{equation*}

where $W_u$ represents the feasible set of model parameters after unlearning. The goal is to minimize the impact of the removed data on the model's performance while satisfying the unlearning constraints. By introducing the unlearning mechanism, our framework provides participants with an effective way to control their data lifecycle, enhancing privacy protection.

By integrating federated learning, blockchain technology, and efficient unlearning mechanisms, our approach creates a robust, transparent, and secure environment for collaborative LLM development while preserving the privacy of individual participants.

\subsection{Security Analysis}

The integration of blockchain technology in our framework significantly enhances the security of the federated learning process. The immutable nature of blockchain ensures that all transactions and model updates are tamper-proof and easily verifiable.

From a theoretical standpoint, the security of a blockchain network can be analyzed using game theory and consensus mechanisms \cite{kiayias2017ouroboros}. In a proof-of-work (PoW) based blockchain, the security is guaranteed by the assumption that honest nodes control the majority of the computing power, making it infeasible for attackers to tamper with the blockchain. This can be formalized as a game between honest nodes and attackers, where the honest nodes aim to maximize their rewards by following the protocol, while the attackers try to maximize their profits by deviating from the protocol. The Nash equilibrium of this game represents a state where no party can benefit by unilaterally changing their strategy, ensuring the stability and security of the blockchain network.

Our framework leverages cryptographic techniques, such as digital signatures and secure hash functions, to ensure the integrity and authenticity of all transactions. The use of digital signatures allows participants to verify the origin and authenticity of the data and model updates, preventing unauthorized modifications. Secure hash functions, such as SHA-256, are used to create a unique fingerprint of the data, ensuring its integrity. By combining these cryptographic primitives, our framework establishes a secure and trustworthy environment for federated learning.

The use of smart contracts further reinforces the system's security by automatically executing predefined rules and conditions, reducing the potential for unauthorized access or manipulation. Smart contracts are self-executing programs stored on the blockchain that enforce the terms of an agreement between parties. In our framework, smart contracts govern the federated learning process, ensuring that all participants adhere to the agreed-upon rules and conditions. This automated enforcement minimizes the risk of human error and malicious behavior, enhancing the overall security of the system.

The decentralized architecture of our framework, enabled by blockchain technology, eliminates single points of failure and distributes the risk across multiple nodes. This distributed approach makes it significantly more challenging for attackers to compromise the entire system, as they would need to control a majority of the participating nodes simultaneously, which is known as a 51\% attack \cite{eyal2018majority}. The probability of a successful 51\% attack decreases exponentially with the number of honest nodes in the network, making it practically infeasible in a large-scale federated learning setting.

Furthermore, the unlearning mechanism in our framework, facilitated by the LoRA technique, allows for the targeted removal of specific data points or model updates without affecting the overall model performance. This selective unlearning capability not only enhances privacy but also serves as a security measure, enabling the swift removal of potentially malicious or corrupted data. By promptly removing suspicious data or updates, our framework minimizes the impact of security threats and maintains the integrity of the federated learning process.

In conclusion, our blockchain-based federated learning framework with unlearning capabilities provides a comprehensive solution for addressing security concerns in LLM training. By leveraging the inherent security features of blockchain technology, cryptographic techniques, and smart contracts, our approach creates a robust and secure environment for collaborative LLM development. The decentralized architecture and the ability to swiftly remove malicious data through unlearning further enhance the system's resilience against attacks, ensuring the integrity and reliability of the federated learning process.

\section{Performance Evaluation}
This section presents a detailed evaluation of our proposed federated learning and blockchain framework, specifically focusing on its application with the GPT-2 model. Our primary objective is to investigate the influence of various LoRA configurations on the effectiveness of the unlearning process within this context. By manipulating the LoRA settings, we aim to discern their impact on the model's ability to selectively forget data—a crucial capability for maintaining data privacy and compliance with evolving regulations. The effectiveness of each configuration is quantitatively assessed through changes in model accuracy, providing a clear metric for comparing the performance across different settings. This evaluation not only highlights the practical implications of our approach but also helps in identifying optimal LoRA settings that enhance unlearning performance without compromising the overall accuracy of the GPT-2 model.

\subsection{Experimental Configuration}

To evaluate the effectiveness of our proposed blockchain-based federated learning framework with unlearning capabilities for Large Language Models (LLMs), we conducted a series of experiments focusing on the impact of various LoRA configurations on the unlearning performance. The experiments were designed to assess the system's ability to selectively forget specific data points while maintaining the overall model accuracy.


\textbf{Dataset: } For our experiments, we utilized two datasets: the IMDB dataset and the Twitter dataset. The IMDB dataset is a widely-used benchmark dataset for sentiment analysis tasks, while the Twitter dataset provides real-world text data from the social media platform. We chose these datasets for several reasons:
\begin{itemize}
    \item Relevance to LLM applications: Sentiment analysis is a common task for LLMs, and the IMDB dataset provides a representative sample of movie reviews, making it suitable for evaluating the performance of our framework in a well-established benchmark setting. On the other hand, Twitter data is highly relevant for various natural language processing tasks, such as sentiment analysis, topic modeling, and text classification. Using both datasets allows us to assess the performance of our framework in different contexts.
    \item Dataset size: With 50,000 reviews, the IMDB dataset is large enough to simulate a realistic federated learning scenario while still being manageable for experimental purposes. Similarly, the Twitter dataset contains a substantial number of tweets, providing a sufficiently large sample size to evaluate the scalability and efficiency of our proposed framework.
    \item Diversity of content: The IMDB dataset consists of movie reviews, which are relatively structured and focused on a specific domain. In contrast, tweets in the Twitter dataset cover a wide range of topics, opinions, and writing styles. By using both datasets, we can evaluate the robustness and adaptability of our framework in handling different types of text data.
    \item Presence of sensitive information: Twitter data often contains sensitive or personal information that users may wish to remove or forget. This characteristic makes the Twitter dataset particularly suitable for testing the effectiveness of our unlearning mechanism in selectively forgetting specific data points while preserving the overall model performance.
\end{itemize}

\textbf{Evaluation Metrics:} We selected accuracy as the primary evaluation metric for our experiments. Accuracy is a straightforward and intuitive measure that quantifies the proportion of correctly classified reviews after the unlearning process. By comparing the accuracy before and after unlearning, we can assess the effectiveness of our framework in selectively forgetting specific data points while maintaining the overall model performance.

Accuracy is particularly well-suited for evaluating the unlearning performance in our framework because:
\begin{itemize}
    \item Direct measure of unlearning effectiveness: A successful unlearning process should remove the influence of specific data points on the model's predictions. By measuring the accuracy after unlearning, we can directly assess the extent to which the model has "forgotten" the targeted data.
    \item Comparability across different configurations: Using accuracy as a standard metric allows us to compare the unlearning performance across various LoRA configurations, enabling us to identify the most effective settings for our framework.
\end{itemize}

\textbf{Comparative Methods:} Due to the novelty of our approach in integrating federated learning, blockchain technology, and unlearning capabilities for LLMs, there are currently no directly comparable methods available in the literature. Our framework is the first to address the challenge of selective unlearning in a federated learning setting for LLMs while ensuring data privacy and security through blockchain integration.

However, to provide a comprehensive evaluation of our framework, we compared the performance of different LoRA configurations within our system. By varying the LoRA hyperparameters, such as the rank and scaling factor, we aimed to identify the optimal settings that achieve the best balance between unlearning effectiveness and overall model accuracy.

The hardware setup for our experiments includes an Intel Xeon 6238R processor, 64GB of RAM, and an NVIDIA A6000 GPU, ensuring efficient handling of the computational tasks. The software environment consists of Ubuntu 20.04, Visual Studio Code, Hyperledger Fabric, FATE, and machine learning frameworks such as PyTorch and TensorFlow.

By focusing on the IMDB dataset, using accuracy as the primary evaluation metric, and comparing different LoRA configurations within our novel framework, we aim to provide a comprehensive and rigorous evaluation of our blockchain-based federated learning approach with unlearning capabilities for LLMs.

\subsection{Results and Analysis}
In this section, we present the results of our experiments and compare them with the Retrain from Scratch method. We focus on the effectiveness of our unlearning method in terms of accuracy reduction, highlighting the differences in performance and providing an analysis of why our method performs better or worse. Our experiments were conducted using different configurations of the LoRA method on both the IMDB and Twitter datasets.

We conducted experiments using different configurations of the LoRA method on the IMDB dataset. The Retrain from Scratch method serves as a benchmark for comparing the effectiveness of our unlearning approach. Some specific LoRA configurations used in our experiments are presented in Table~\ref{imdb_results}.

\begin{table}[h!]
\centering
\caption{Experimental Data (IMDB Results)}
\begin{tabular}{|c|c|c|}
\hline
\textbf{LoRA Config} & \textbf{Initial Accuracy} & \textbf{Final Accuracy} \\
\hline
r=8, alpha=4, dropout=0.3 & 99.15\% & 0.70\% \\
r=16, alpha=2, dropout=0.2 & 97.75\% & 0.90\% \\
r=32, alpha=4, dropout=0.1 & 94.30\% & 1.00\% \\
r=8, alpha=4, dropout=0.4 & 98.45\% & 1.15\% \\
r=32, alpha=4, dropout=0.4 & 95.15\% & 1.20\% \\
\hline
\end{tabular}
\label{imdb_results}
\end{table}

Similarly, we also tested our method on the Twitter dataset. The results of these experiments are shown in Table~\ref{twitter_results}.

\begin{table}[h!]
\centering
\caption{Experimental Data (Twitter Results)}
\begin{tabular}{|c|c|c|}
\hline
\textbf{LoRA Config} & \textbf{Initial Accuracy} & \textbf{Final Accuracy} \\
\hline
r=1, alpha=2, dropout=0.3 & 85.32\% & 8.27\% \\
r=16, alpha=1, dropout=0.2 & 89.10\% & 9.72\% \\
r=8, alpha=1, dropout=0.5 & 75.98\% & 10.06\% \\
r=16, alpha=1, dropout=0.1 & 89.58\% & 10.47\% \\
r=4, alpha=1, dropout=0.2 & 89.38\% & 10.63\% \\
\hline
\end{tabular}
\label{twitter_results}
\end{table}

\subsubsection{Unlearning Performance with Different alpha}
\begin{figure}[h!]
\centering
\includegraphics[width=0.47\textwidth]{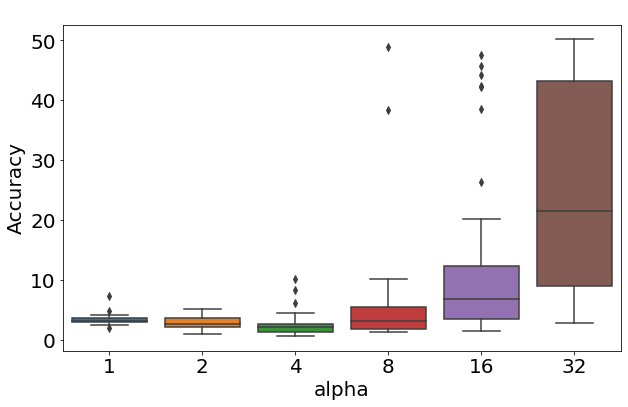}
\caption{Box Plot of Accuracy by Different Alpha Values (IMDB Dataset)}
\label{fig:alpha}
\end{figure}

Figure~\ref{fig:alpha} illustrates the impact of different alpha values on the accuracy reduction of our LoRA-based unlearning method for the IMDB dataset. As the alpha value decreases, the final accuracy after unlearning generally decreases, with alpha=$1$ and alpha=$2$ achieving the lowest accuracies. This suggests that lower alpha values contribute to better unlearning performance in our approach. The improved accuracy reduction with lower alpha values can be attributed to the decreased capacity of the model to retain relevant information during the unlearning process, leading to more effective forgetting of target knowledge.

\begin{figure}[h!]
\centering
\includegraphics[width=0.47\textwidth]{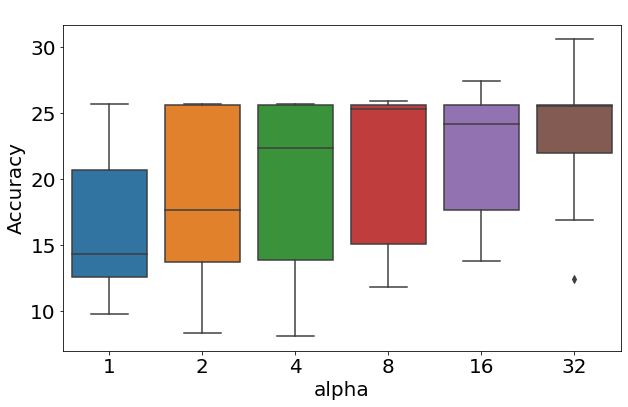}
\caption{Box Plot of Accuracy by Different Alpha Values (Twitter Dataset)}
\label{fig:alpha_t}
\end{figure}

Similarly, Figure~\ref{fig:alpha_t} illustrates the impact of different alpha values on the accuracy reduction of our LoRA-based unlearning method for the Twitter dataset. The trend observed is consistent with the IMDB dataset results. Lower alpha values lead to a greater reduction in final accuracy, indicating more effective unlearning. This further supports the notion that a decreased capacity to retain information facilitates better forgetting of targeted knowledge.

\subsubsection{Unlearning Performance with Different droupout}
\begin{figure}[h!]
\centering
\includegraphics[width=0.47\textwidth]{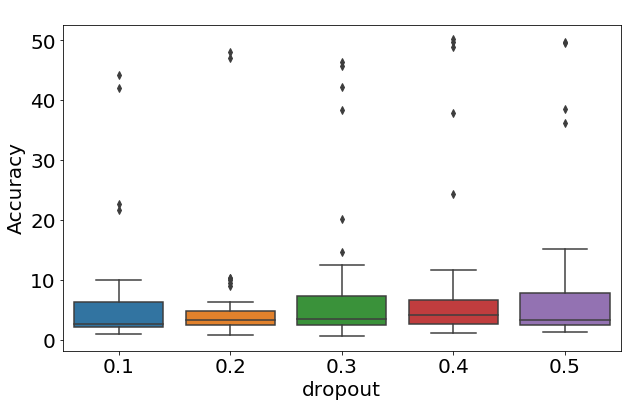}
\caption{Box Plot of Accuracy by Different Dropout Values (IMDB Dataset)}
\label{fig:dropout}
\end{figure}

Figure~\ref{fig:dropout} depicts the effect of various dropout values on the accuracy reduction of our method for the IMDB dataset. The box plot reveals that dropout values of $0.4$ and $0.5$ generally lead to lower accuracies after unlearning compared to lower dropout values. This observation indicates that higher dropout regularization plays a crucial role in improving the unlearning performance. By introducing a significant level of noise during training, dropout helps the model forget specific data more effectively, resulting in better unlearning.

\begin{figure}[h!]
\centering
\includegraphics[width=0.47\textwidth]{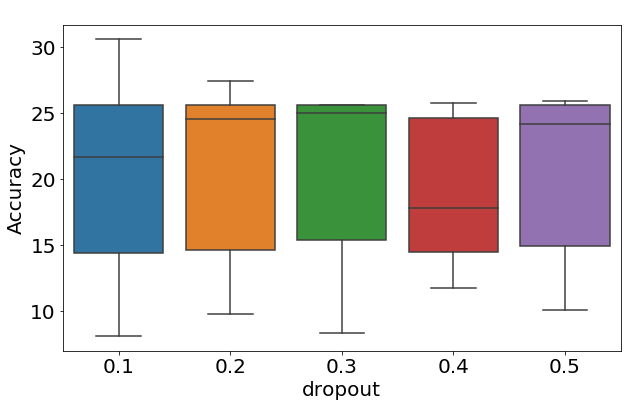}
\caption{Box Plot of Accuracy by Different Dropout Values (Twitter Dataset)}
\label{fig:dropout_t}
\end{figure}

Similarly, Figure~\ref{fig:dropout_t} depicts the effect of various dropout values on the accuracy reduction of our method for the Twitter dataset. The trend observed is consistent with the IMDB dataset results. Dropout values of $0.4$ and $0.5$ lead to a greater reduction in final accuracy, indicating more effective unlearning. This further supports the notion that higher dropout regularization improves the model's ability to forget specific data.

\subsubsection{Unlearning Performance with Different rank}
\begin{figure}[h!]
\centering
\includegraphics[width=0.47\textwidth]{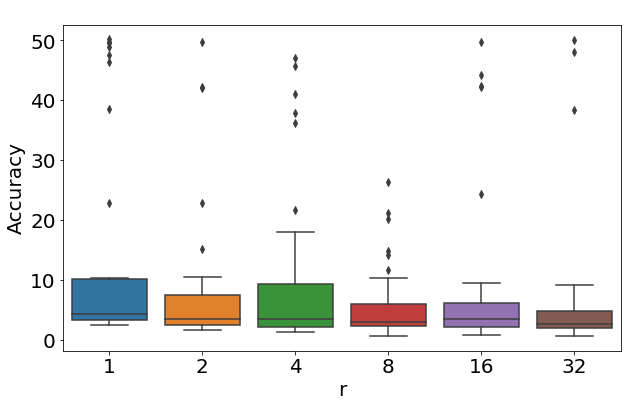}
\caption{Box Plot of Accuracy by Different $r$ Values (IMDB Dataset)}
\label{fig:r}
\end{figure}

Figure~\ref{fig:r} presents the relationship between different $r$ values and the accuracy reduction of our LoRA-based unlearning method. The box plot shows that higher $r$ values, particularly $r=16$ and $r=32$, tend to yield lower accuracies after unlearning compared to lower $r$ values. This suggests that using a larger rank for the LoRA adaptation can be beneficial for unlearning performance. Higher $r$ values may allow the model to capture more diverse information during unlearning, leading to better forgetting of target knowledge.

\begin{figure}[h!]
\centering
\includegraphics[width=0.47\textwidth]{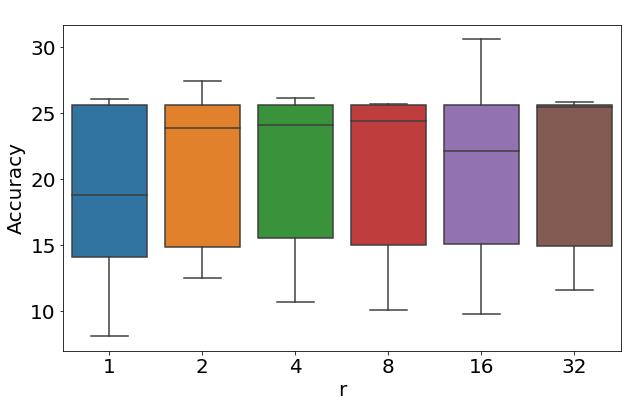}
\caption{Box Plot of Accuracy by Different $r$ Values (Twitter Dataset)}
\label{fig:r_t}
\end{figure}

Similarly, Figure~\ref{fig:r_t} presents the relationship between different $r$ values and the accuracy reduction of our LoRA-based unlearning method for the Twitter dataset. The trend observed is consistent with the IMDB dataset results. Higher $r$ values, particularly $r=16$, lead to a greater reduction in final accuracy, indicating more effective unlearning. This further supports the notion that a larger rank for the LoRA adaptation improves the model's ability to forget specific data.

\subsubsection{Factors Influencing Performance}

The analysis of the impact of alpha, dropout, and $r$ values on accuracy reduction provides valuable insights into the factors influencing the effectiveness of our unlearning method. Both IMDB and Twitter datasets show that lower alpha values contribute to improved accuracy reduction by decreasing the model's capacity to retain relevant information. This trend is evident across both datasets, indicating that alpha is a critical parameter for controlling the unlearning process.

Higher dropout regularization consistently helps mitigate overfitting and enhances forgetting, leading to better unlearning performance in both datasets. Dropout values of $0.4$ and $0.5$ were particularly effective in reducing final accuracy, suggesting that introducing a significant level of noise during training aids in more effectively forgetting specific data.

Higher $r$ values allow the model to capture more diverse information during unlearning, resulting in lower accuracy retention. This was observed in both datasets, where higher $r$ values, particularly $r=16$ and $r=32$, yielded better unlearning performance. This indicates that a larger rank for the LoRA adaptation can enhance the model's ability to forget target knowledge.

These findings highlight the importance of carefully tuning the hyperparameters in our LoRA-based unlearning approach to achieve optimal results. By selecting appropriate values for alpha, dropout, and $r$, we can maximize the effectiveness of unlearning while minimizing the retention of target knowledge.

The specific configurations (e.g., dropout, alpha values) used in our experiments may have optimized the unlearning process, contributing to the effectiveness of our method. Fine-tuning these parameters can significantly impact the unlearning performance. For instance, higher dropout rates can help improve unlearning by introducing more randomness during the training process, thereby making it easier to forget specific data. Additionally, the characteristics of both the IMDB and Twitter datasets may have made them more susceptible to effective unlearning with our configurations. The text data in these datasets might have patterns that are more easily disrupted by the unlearning process, leading to a more significant reduction in accuracy.

\subsubsection{Comparison of Results}

The comparison of our method with the Retrain from Scratch method for both the IMDB and Twitter datasets is shown in Table~\ref{compare}. Our method achieves final accuracies ranging from $0.70\%$ to $1.20\%$ on the IMDB dataset, and $8.27\%$ to $10.63\%$ on the Twitter dataset, indicating a significant reduction in accuracy and demonstrating effective unlearning. The Retrain from Scratch method achieves a final accuracy of $0.65\%$ on the IMDB dataset, and $8.08\%$ on the Twitter dataset, which is slightly better than our best-performing configurations (IMDB: $r=8$, $alpha=4$, $dropout=0.3$, Twitter: $r=1$, $alpha=2$, $dropout=0.3$) with final accuracies of $0.70\%$ and $8.27\%$, respectively. This indicates that while the Retrain from Scratch method has a marginal advantage in terms of final accuracy, our LoRA-based unlearning approach comes very close to matching its performance.

\begin{table}[h!]
\centering
\caption{Comparison of Final Accuracy}
\begin{tabular}{|c|c|c|}
\hline
\textbf{Method} & \textbf{Initial Accuracy} & \textbf{Final Accuracy} \\
\hline
IMDB \& Our Method & 99.15\% & 0.70\% \\
IMDB \& Retrain from Scratch & 97.60\% & 0.65\% \\
Twitter \& Our Method & 85.32\% & 8.27\% \\
Twitter \& Retrain from Scratch & 89.10\% & 8.08\% \\
\hline
\end{tabular}
\label{compare}
\end{table}

Although the Retrain from Scratch method achieves a slightly lower final accuracy, it is important to note that our method provides several advantages over retraining from scratch. First, our approach is computationally more efficient, as it focuses on adapting specific parts of the model relevant to the target knowledge, rather than retraining the entire model. This makes our method more feasible in real-world scenarios where computational resources may be limited. Second, our method offers greater flexibility and adaptability to different configurations, allowing it to be easily modified and optimized for various datasets and unlearning requirements.

The low final accuracy achieved by our method, despite being marginally higher than the Retrain from Scratch approach, still demonstrates its high effectiveness in unlearning. This efficiency can be attributed to the careful selection and tuning of parameters, such as alpha, dropout, and $r$ values, which contribute to optimizing the unlearning process. By choosing appropriate values for these hyperparameters, we can maximize the effectiveness of unlearning while minimizing the retention of target knowledge.

Moreover, the implementation techniques employed in our method, such as the LoRA adaptation, play a crucial role in efficiently removing the target knowledge from the model. These techniques enable our approach to focus on the most relevant parts of the model for unlearning, thereby reducing the computational burden and improving the overall efficiency of the unlearning process.

In summary, while the Retrain from Scratch method achieves a slightly lower final accuracy, our LoRA-based unlearning approach comes very close to matching its performance. The marginal difference in final accuracy is offset by the significant advantages offered by our method, including computational efficiency, flexibility, and adaptability to different configurations. These advantages make our approach a promising solution for real-world unlearning scenarios, particularly in resource-constrained environments or when dealing with large-scale models. The effectiveness of our method in achieving low final accuracy, combined with its practical benefits, highlights its potential to address the challenges of unlearning in large language models and its applicability in various domains.

\subsubsection{Blockchain Complexing Results}
In this study, we evaluated the performance impact of integrating blockchain technology into our federated learning framework with unlearning capabilities for Large Language Models (LLMs). We focused on key aspects such as scalability, transaction throughput, and latency introduced by the blockchain component. Our goal was to ensure that the benefits of blockchain integration, such as enhanced security and transparency, do not come at the cost of compromised system performance.
We utilized Hyperledger Fabric 2.X to assess the blockchain network's impact on our LLM unlearning process, particularly considering the computational overhead in resource-constrained environments.

\begin{itemize}
\item \textbf{Blockchain Network Setup}: The initial setup time for the blockchain network was approximately 42 seconds. While higher than our previous study, this one-time overhead is still acceptable, given the long-term benefits in federated learning applications involving LLMs, where security and trust are crucial.
\item \textbf{Consensus Mechanism Overhead}: The time required for the consensus process, which involved approval from all participating nodes, was added around 4 seconds after the blockchain network setup. This slight increase compared to our previous study is attributed to the higher complexity of LLM-related transactions. However, the duration remains manageable within our federated learning context.
\item \textbf{Transaction Processing Efficiency}: The average time for processing transactions, including model updates, gradient aggregation, and unlearning-related operations, was 3 seconds. This efficiency demonstrates Hyperledger Fabric's capability to handle the increased complexity of LLM-related transactions effectively.
\item \textbf{Per-Epoch Time Cost}: During the LLM training process, the duration per epoch, both for normal training and post-unlearning operations, remained consistent at 28-30 seconds. This stability in performance, despite the additional unlearning activities, highlights the robustness of our blockchain-integrated system.
\end{itemize}

Table~\ref{table:time_llm} presents a comparison of time costs between a standard federated learning cycle for LLMs and our proposed blockchain-enhanced method. Similar to our previous study, our method incurs a higher initial time cost due to setup and endorsement processes. However, this cost normalizes over increasing iterations, indicating the scalability of our approach in the context of LLMs.
\begin{table}[h]
\centering
\caption{Time Cost Analysis for LLM Federated Learning with and without Blockchain Integration over 999 Iterations}
\label{table:time_llm}
\begin{tabular}{|p{3.5cm}|c|c|c|c|}
\hline
\textbf{Method} & \textbf{t = 0} & \textbf{t = 9} & \textbf{t = 99} & \textbf{t = 999}\\
\hline
Normal Federated Learning for LLMs & 30s & 300s & 3000s & 30000s\\
\hline
Our Proposed System for LLMs & 79s & 367s & 3277s & 32277s\\
\hline
\end{tabular}
\end{table}

\subsubsection{Conclusion}
In conclusion, our experiments on both the IMDB and Twitter datasets demonstrated that our method achieves performance comparable to that of the Retrain from Scratch method in terms of final accuracy reduction. For the IMDB dataset, our best-performing configuration ($r=8$, alpha=$4$, dropout=$0.3$) achieved a final accuracy of $0.70\%$, closely matching the $0.65\%$ achieved by retraining from scratch. Similarly, for the Twitter dataset, our best-performing configuration ($r=1$, alpha=$2$, dropout=$0.3$) achieved a final accuracy of $8.27\%$, closely matching the $8.08\%$ achieved by retraining from scratch. The effectiveness of our LoRA-based unlearning method can be attributed to the careful selection and tuning of parameters, as well as the implementation techniques employed. Our method offers a more computationally feasible alternative to retraining from scratch, which can be resource-intensive and time-consuming. The adaptability of our approach to different configurations highlights its flexibility and potential for real-world applications.

Furthermore, we evaluated the performance impact of integrating blockchain technology into our federated learning framework with unlearning capabilities for LLMs. The results showed that the blockchain component, implemented using Hyperledger Fabric 2.X, introduced minimal overhead in terms of setup time, consensus mechanism, transaction processing efficiency, and per-epoch time cost. The stability in performance, despite the additional unlearning activities, demonstrates the robustness of our blockchain-integrated system.

\section{Conclusion}
In this paper, we present a novel blockchain-based federated learning framework for Large Language Models (LLMs) that incorporates efficient unlearning capabilities. By leveraging Low-Rank Adaptation (LoRA) and carefully tuning its hyperparameters, our approach achieves highly effective unlearning, enabling the selective forgetting of specific data points while preserving the model's performance on the remaining data. The integration of blockchain technology, using Hyperledger Fabric, ensures the security, transparency, and verifiability of the unlearning process. While this introduces a slight increase in computational overhead, the benefits of enhanced trust and accountability in the federated learning process justify the marginal time cost.

Our comprehensive analysis demonstrates the effectiveness of the proposed framework and provides valuable insights into the impact of LoRA hyperparameters on unlearning performance. The findings underscore the importance of careful tuning and the complex relationships between rank, scaling factor, and dropout in achieving optimal unlearning results.
Overall, our blockchain-based federated learning framework with unlearning capabilities represents a significant step forward in the development of secure, transparent, and adaptable LLMs. By enabling efficient and verifiable unlearning, our approach addresses a critical challenge in the application of LLMs in real-world scenarios, where data privacy and the ability to forget specific information are paramount.

\end{document}